\documentclass[twocolumn,showpacs,amsmath,amstex,amssymb,mathfonts,prl]{revtex4}
\usepackage{graphicx,bm,units}
\usepackage{epstopdf}
\begin{document}
\title{Singular Behavior of Eigenstates in Anderson's Model of Localization}

\author{S. Johri$^{1}$ and R. N. Bhatt$^{1,2}$}
\affiliation{$^1$ Department of Electrical Engineering, Princeton University, Princeton, NJ 08544}
\affiliation{$^2$ Princeton Center for Theoretical Science, Princeton University, Princeton, NJ 08544}

\begin{abstract}
We observe a singularity in the electronic properties of the Anderson Model of Localization with bounded diagonal
disorder, which is clearly distinct from the well-established mobility edge (localization-delocalization transition) that occurs in dimensions $d>2$. We present results of numerical calculations for Anderson's original (box) distribution of onsite disorder in dimensions $d$ = $1$, $2$ and $3$. To establish this hitherto unreported behavior, and to understand its evolution with disorder, we contrast the behavior of two different measures of the localization length of the electronic wavefunctions - the averaged inverse participation ratio and the Lyapunov exponent. Our data suggest that Anderson's model exhibits richer behavior than has been established so far.
\end{abstract}


\date{\today}
\maketitle

The Anderson model of localization, formulated over fifty years ago, has been the foundation on which our understanding of the effects of disorder on electronic systems has been built. In his original paper\cite{anderson}, Anderson demonstrated the existence of localized states in lattice systems in three dimensions with sufficient disorder. Following Anderson's work, Mott and Twose\cite{mott} showed that in one dimension electron states become localized with arbitrary small disorder. It took, however, over twenty years to establish the situation for two dimensions, with the seminal paper on the scaling theory of localization\cite{abrahams}, which concluded that all states were localized there also, but with exponentially large localization lengths for weak disorder. Concurrently, the work of Wegner\cite{wegner}  mapping the localization problem on the sigma model opened the doors for a systematic study of  the localization transition in $2+\epsilon$  dimensions for models with disorder belonging to different symmetry classes{\cite{evers}. Sophisticated numerical techniques were developed to address the problem of localization\cite{evers},\cite{kramer},\cite{slevin}, and study not only critical exponents, but also multifractality of the wavefunctions, as well as eigenvalue statistics at the critical point. A 50-year commemorative volume has recently appeared\cite{50years}.
\newline
\hspace*{10 pt}
Following Anderson's paper, there was concern that the mobility edge, signaling the transition from localized to extended states may be accompanied by a singularity of the density of states ($DOS$). However, a pioneering work by Edwards and Thouless\cite{thouless} proved that the $DOS$ in the original Anderson model (with a uniform but bounded distribution of disorder) was analytic in a large region near the band center for any value of the disorder, and thus, by implication, there was no singularity at the Anderson transition at these energies. Wegner\cite{wegner} showed that for Gaussian disorder, there was no singularity in the $DOS$ at any energy for any nonzero disorder. Since the localization transition was present with Gaussian disorder as well, this laid to rest any speculation about the occurrence of any singular behavior of the $DOS$ at the localization transition.

The lack of singular behavior in the problem of Anderson localization has given rise to the popular belief that apart from the divergence of the localization length and related behavior (e.g. multifractality) at the transition, properties are generally smooth everywhere. However, for the 1D Anderson model, it was shown\cite{derrida} showed that the weak disorder expansion of the Lyapunov exponent is non-analytic at band center and other commensurate fillings; a comprehensive study has recently been done\cite{kravtsov}. Here we provide evidence of a rather prominent singularity of a different kind; in particular, we show that for Anderson's original model, outside the bounds of Edwards and Thouless\cite{thouless}, there is a singular behavior of electronic eigenstates, as measured by the ensemble averaged inverse participation ratio, and possibly of the $DOS$ itself. We find this behavior in the localized region of the phase diagram, in one, two and three dimensions, and expect it to be present in higher dimensions as well. Furthermore, we find singular behavior for all bounded distributions of on-site disorder (only) that we have considered, but not for unbounded distributions like Gaussian disorder, the latter being consistent with the result of an analytic $DOS$ for Gaussian disorder\cite{wegner}. 

We recall the single-band tight-binding Hamiltonian studied by Anderson\cite{anderson} -
\vspace{-10pt}
\begin{equation}\label{eq:ham}
H = \sum_{i}\epsilon_i |i><i| + V \sum_{i,j} |i><j|						
\vspace{-10pt}
\end{equation}
where $|i>$ denotes a state localized at the site $i$, the sum $i$ is over sites of a $d$-dimensional hypercubic lattice,  and $i,j$ are nearest neighbors. Further, $\epsilon_i$ are independent random variables distributed uniformly in the (bounded) interval $(-W/2, W/2)$, i.e. $P(\epsilon_i) = 1/W$ in the aforesaid interval, and $P(\epsilon_i) = 0$ outside. For such a distribution, the $DOS$ is nonzero only in the bounded interval $(- E_B, E_B)$, where  $E_B = W/2 + ZV$ is the true band edge and $Z = 2d$ is the coordination number for hypercubic lattices. [Quite generally, for  $P(\epsilon_i)$ symmetric around a mean (which can be taken to be zero without loss of generality), the $DOS$ is also symmetric for bipartite lattices like the hypercubic lattice]. For this model, the $DOS$ was shown to be analytic for $|E| < W/2 - ZV$\cite{thouless}. {\it In what follows, we use $ V = 1 $, as our unit of energy.}

Two canonical quantities used to study localization numerically are (i) the inverse participation ratio, and (ii) the Lyapunov exponent. The inverse participation ratio of any wavefunction $\Psi = \sum_{i} {a_i |i>}$is defined as:
\vspace{-10pt}
\begin{equation}\label{eq:ipr}
IPR_{\Psi} =	\frac{\sum_{i} |a_i|^4}{(\sum_{i} |a_i|^2)^2}		
\vspace{-10pt}				
\end{equation}

To determine the $IPR_{\Psi}$, we diagonalize the Hamiltonian (\ref{eq:ham}) for lattices of size $L^d$ (in $d$-dimensions), for a specified value of $W/V$, and compute the inverse participation ratio for each eigenstate $\Psi$. We then compute the ensemble averaged $IPR$ for eigenstates with energies in a small interval around any given energy $E$, by collecting data for as many samples as are needed for $IPR(E)$ to converge, and plot $IPR(E)$ versus $E$. For energies $E$ within the localized regime, it is easy to see that $IPR$ is inversely proportional to the number of sites the typical wavefunction resides on and therefore $IPR$ reaches a constant value as the size of the system $L\rightarrow \infty$. For $E$ within extended states, $IPR$ is expected to decay as $L^{-d}$, in $d$ dimensions, while at the mobility edge, $IPR$ decays as a non-trivial power law, related to the multifractal nature of the eigenstates at the critical energy\cite{evers}.

	The Lyapunov exponent characterizing eigenstates at a give energy $E$, $Ly(E)$, is determined using a quasi-one-dimensional structure, with a fixed width $M$ in the transverse dimensions (i.e. a cross section of $M^{d-1}$) and a (much longer) length $L$ in one dimension, which is allowed to become as large as necessary to obtain the $L \rightarrow \infty$ limit. $Ly(E)$ gives the exponential decay of the wavefunctions at energy E (quasi-one-dimensional), and is thus inversely related to the localization length  $\xi_M (E)$ at $E$. (To obtain the true behavior of the localization length in the thermodynamic limit for a $d > 1$ dimensional system, the limit $M \rightarrow \infty$ has to be dealt with; in this paper, we restrict ourselves to $Ly(E)$ for $d = 1$, where no such extrapolation is required). In $d=1$, $Ly(E)$ is obtained\cite{kramer} from the eigenvalues of the transfer matrix (which come in inverse pairs), for each value of the disorder parameter $W/V$, by increasing the size of the system $L$ till convergence is reached.

\begin{figure}
	\centering
	\vspace{-15 pt}
		\hspace*{-20pt}
	\includegraphics[width=9cm,keepaspectratio]{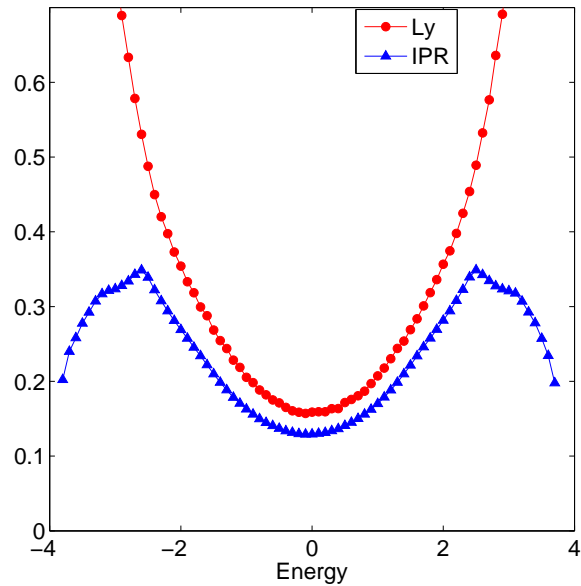}
	\caption{$IPR(E)$ and $Ly(E)$ for $W=4$ for the 1D Anderson model with ``box'' distribution}
	\label{fig:1}
\end{figure}

	In $d = 1$, where all states are known to be localized for all nonzero disorder ($W$), both $IPR(E)$ and $Ly (E)$ are expected to be inversely proportional to the localization length $\xi(E)$. In Figure 1, we plot both $IPR(E)$ and $Ly(E)$ as a function of $E$ for a typical value $W = 4$. (For both plots, the size of the system has been taken to be large enough for the quantities to converge within our statistical errors). As can be seen, the two quantities track each other reasonably well in the middle of the band, both increasing as one moves away from the band center (implying a decreasing localization length, as would be expected). However, as one approaches the band edges, $Ly(E)$ continues to rise, whereas $IPR(E)$ changes course, and appears to go to zero at the band edge. Furthermore, upon close examination, the downturn appears to be accompanied by a non-analyticity at the maximum value of the $IPR$.

	The non-monotonic behavior of $IPR(E)$ is easily understood in terms of Lifshitz states \cite{kramer}, \cite{lifshitz}. The states at the very edge of the band are due to rare configurations of a cluster of contiguous sites which all have an on-site energy close to $W/2$ (or $-W/2$); the larger the cluster, the closer the cluster eigenstate energy ($E_{CL}$) to the band edge, $E_B$ ($(E_B - E_{CL}) \propto 1/L^2$, where $L$ is the linear dimension of the cluster; such a result follows from a particle-in-a-box like considerations\cite{kramer}). In the case of Lifshitz states, the eigenstates actually spread over a larger number of sites as the band edge is approached; nevertheless, their exponential decay at long distances keeps getting faster. Consequently, $Ly(E)$ grows monotonically, while $IPR(E)$ decreases as the band edge is approached, going to zero like $|E_B - E|^{d/2}$.

	Since rare clusters giving rise to the tail of the Lifshitz states occur with probability $exp[-cL^d]$ for clusters of linear dimension $L$, one can show {\cite{lifshitz}} that near the band-edge, the electronic density of states, $N(E) \propto exp [-C/|E_B - E|^{-d/2}]$, where $c$ and $C$ are constants. Thus, while much is known\cite{lifshitz} about the behavior asymptotically near the band edge, it is also tempered by the fact that the $DOS$ goes exponentially fast to zero at the band-edge with an essential singularity; further, this exponential drop becomes more pronounced in higher dimensions, so as to be of less practical significance. 

	However, the phenomenon of resonant tunneling which gives rise to the Lifshitz tail, is not limited to large clusters. In fact, in the large disorder limit ($W >> V$), resonance between a single pair of sites gives rise to states that are outside the disorder bandwidth $W$, by $O(V)$, which is parametrically larger than the perturbation series for typical ``Anderson-localized'' states, for which the energy shift in a locator expansion\cite{wolfle} is $V^2/W$. States residing on two (or a small number of) resonant states are not exponentially rare, but only down by a power-law factor in the expansion parameter ($V/W$) at large disorder. Such local resonance can give rise to special effects, such as extended states in dimer models\cite{phillips}, and on tree structures\cite{aizenman} and, as we show below, also for the Anderson model of Eq. (1) on hypercubic lattices.

\begin{figure}
	\vspace{-10pt} 
	\hspace*{-20pt}	
	\includegraphics[width=10cm,keepaspectratio]{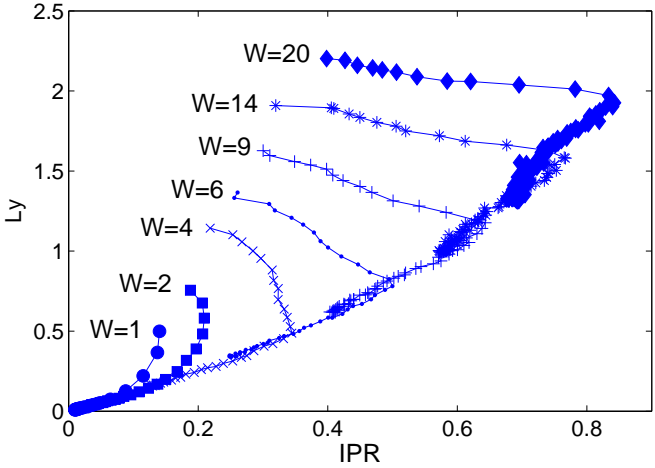}
		\vspace{-22pt}
	\caption{$Ly(E)$ vs $IPR(E)$ for different values of disorder $W$ for the 1D Anderson model.}
	\label{fig:2}
\end{figure}

	Figure (2) plots $Ly(E)$ versus $IPR(E)$ for different values of $W/V$ for the 1D Anderson model. As can be seen, the implicit plots show two clear branches for $W/V = 4$ and greater, confirming the non-analytic behavior of $IPR(E)$ at its maximum. For smaller $W/V$, the two-branched curves lose the sharp bend and appear to become analytic, but remain re-entrant (bending backwards); as $W\rightarrow 0$, the re-entrant behavior disappears as well.
\begin{figure}
	\centering
	\vspace{-5pt} 
		\includegraphics[width=8.5 cm,keepaspectratio]{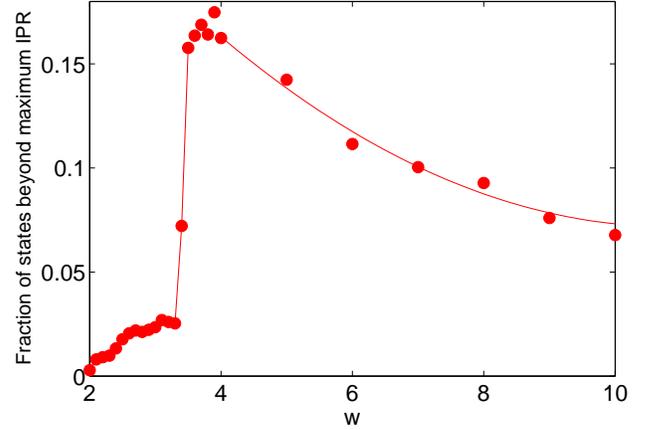}
	\vspace{-15pt}
	\caption{Abrupt rise in fraction of ``resonant'' states near $W=3.8$ for the 1D Anderson model.}
	
	\label{fig:3}
\end{figure}		
	For each value of $W/V$, we divide the states into the two branches, separated by the maximum value of $IPR(E)$. {\it For the positively/negatively correlated branches, we call the states ``typical'' - Anderson localized, and ``resonant'' - Lifshitz like, states respectively}. By determining the total number of states in each branch we plot the fraction of ``resonant'' states ({\it as defined above}) of the total number of states in Figure 3 as a function of the disorder parameter $W/V$. As can be seen, this kind of division suggests that there is a singular increase in the effective number of resonant states around $W/V = 3.8$, to a value of $\approx 17\%$, a fraction large enough that they cannot reasonably be called ``rare-fluctuation'' effects, as Lifshitz tail states typically are.
	
\begin{figure}
	\centering
	\vspace{30pt} 
	\hspace*{-5 pt}
		\includegraphics[width=8.5 cm,keepaspectratio]{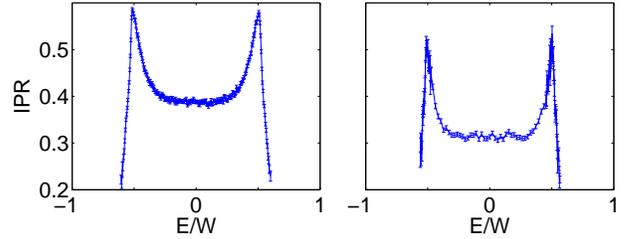}
		\vspace{-12pt}
	\caption{$IPR$ of the 2D Anderson model at $W=20$ (left) and a 3D Anderson model at $W=32$ (right)}
	\label{fig:4}
\end{figure}
The above singular behavior is seen in one dimension for which all states are localized for nonzero W. Clearly, therefore, the existence of this phenomenon should not depend on dimensionality; in fact, we see the same behavior for $IPR(E)$ for the Anderson model with the box distribution of disorder in two and three dimensions (in the localized phase) also (Fig 4).

It is instructive to look at the whole distribution of the $IPR$ at different energies in the band instead of just its average value $IPR(E)$ at each energy. In Fig. 5, we plot this distribution (in $d = 1$) for large disorder ($W = 10$), when most eigenstates are very strongly localized, and the distribution can be easily interpreted. The $IPR$ has a bimodal distribution with two peaks around 0.5 and 0.9, indicating that most of the states have large amplitudes on either 2 or 1 sites respectively. The distribution does not change much as we move from the center of the band towards the edges, till at a certain value of energy, we suddenly lose all the 1-site i.e. Anderson-type wavefunctions. This is exactly the energy at which the sharp downturn of the $IPR$ is observed.

\begin{figure}
	\centering
	\vspace{-5pt}
	\hspace*{-10pt}
		\includegraphics[width=9cm,keepaspectratio]{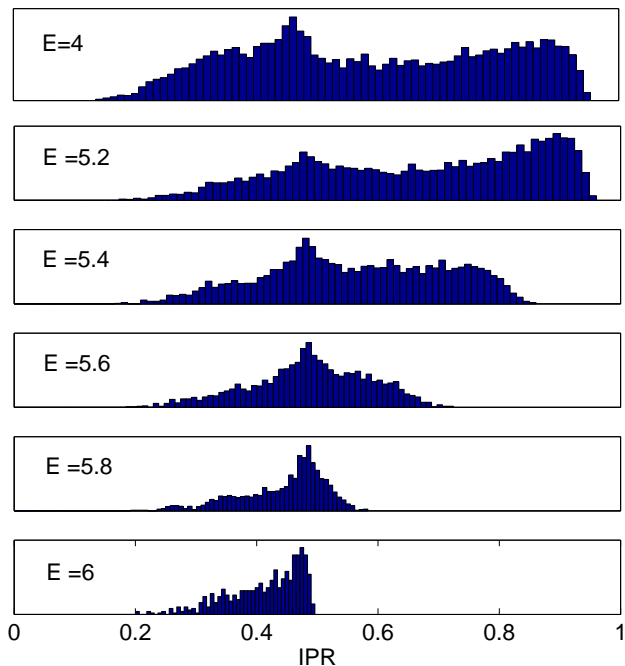}
		\vspace{-25pt}
	\caption{Distribution of $IPR$ at different energies in the band for $W=10$ for the 1D Anderson model.}
	\label{fig:5}
\end{figure}

We have also explored other bounded distributions $P(\epsilon_i)$, and find  a similar sharp transition as a function of energy $E$ between typical Anderson-localized states, and resonant states. A detailed analysis of these results will be presented elsewhere\cite{johri-bhatt1}. However,  no such singularity is observed numerically when the onsite energies are Gaussian distributed, even when we reach very low values of DOS (down to $.01\%$ of the peak). We believe the lack of singular behavior is due to the fact that the Gaussian distribution is unbounded; and our finding is consistent with Wegner's result (no singularity in $DOS$ for Gaussian disorder\cite{wegner}).

As the singularity is seen only for bounded disorder, it is appropriate to ask whether such models are realistic for actual experimental systems. While the naive guess would favor unbounded distributions, for real disordered systems, chemical considerations (e.g. electron affinities or ionization potentials of locally stable random clusters) would in fact suggest that models with bounded disorder are closer to reality. Sharp thresholds have been seen in optical absorption\cite{capizzi} and $DOS$\cite{bhatt-rice} of dopant clusters in the positionally disordered system of doped semiconductors. For the traditional alloy model with a bimodal distribution of onsite energies, the situation is even more severe, with several critical energies separating eigenstates of different types\cite{johri-bhatt1}, especially in the large disorder limit. 

Since the singularity we find persists (and is most evident) for large disorder (i.e., large $W$), and a majority of eigenstates are localized mainly on one or two sites in that limit (see Fig. 5), we have solved a simple two-site Anderson model with a uniform (box) distribution for the onsite energies, for which we have been able to derive analytical expressions for both the $DOS$ and $IPR(E)$. We find that for such a toy model\cite{johri-bhatt2}, there is singular behavior of both quantities at a critical energy $ |E| = \sqrt{(W/2)^2+V^2} $, which is in good agreement with the critical energy determined numerically for the thermodynamic limit for large $W$. We have also numerically computed both quantities for finite one-dimensional rings of lattice sites, and find that while the singularity in the $DOS$ gets weaker with size, that for $IPR(E)$ actually gets stronger with increasing size, consistent with there being a genuine singularity in the thermodynamic limit. The behavior of the $DOS$ in the thermodynamic limit will be reported elsewhere\cite{johri-bhatt1}.

In conclusion, we have numerically analyzed the behavior of the eigenstates of Anderson's original model for single-particle localization on hypercubic lattices in dimensions $ d = 1, 2$ and $3 $ with on-site disorder. By focusing on two measures of the localization length - the inverse participation ratio ($IPR$) and the Lyapunov exponent in the localized phase, we find that the two measures show distinct behavior as one moves from the band center towards the band edge. This divergence of behavior is accompanied by a singularity of the $IPR$ at a critical energy that separates typical Anderson-localized states from resonant states. Higher moments of the electronic wavefunctions also display singular behavior at the same energy. This critical energy is found for bounded disorder distributions (but not for unbounded distributions), and is distinct both from the mobility edge and the band edge. Possible experimental consequences of such an abrupt change in behavior are being investigated.

This work was supported by DOE grant DE-SC20002140. RNB acknowledges hospitality of the Aspen Center for Physics.


\end{document}